\documentclass[aps,prl,reprint,superscriptaddress,twocolumn]{revtex4-1}

\usepackage{amsfonts}
\usepackage{amsmath}
\usepackage{ulem}
\usepackage{bm}
\usepackage{graphicx}
\usepackage[caption=false]{subfig}
\usepackage{lipsum}
\usepackage{array,multirow}
\usepackage[utf8x]{inputenc}
\newcommand{\ket}[1]{\vert#1\rangle}

\def\opone{\leavevmode\hbox{\small1\kern-3.8pt\normalsize1}}
\usepackage{color}

\begin{document}
	
	\title{Improved Light-Matter Interaction for Storage of Quantum States of Light in a Thulium-Doped Crystal Cavity\\}
	
	\author{Jacob H. Davidson}
	\affiliation{Institute for Quantum Science and Technology, and Department of Physics \& Astronomy, University of Calgary, 2500 University Drive NW, Calgary, Alberta, T2N 1N4, Canada}
	\affiliation{QuTech and Kavli Institute of Nanoscience, Delft University of Technology, 2600 GA Delft, The Netherlands}
		\author{Pascal Lefebvre}
	\affiliation{Institute for Quantum Science and Technology, and Department of Physics \& Astronomy, University of Calgary, 2500 University Drive NW, Calgary, Alberta, T2N 1N4, Canada}
		\author{Jun Zhang}
	\affiliation{Institute for Quantum Science and Technology, and Department of Physics \& Astronomy, University of Calgary, 2500 University Drive NW, Calgary, Alberta, T2N 1N4, Canada}
	\author{Daniel Oblak}
	\altaffiliation{Available for correspondence and requests for materials addressed to D. Oblak (email: \mbox{d.oblak@ucalgary.ca).}}
	\affiliation{Institute for Quantum Science and Technology, and Department of Physics \& Astronomy, University of Calgary, 2500 University Drive NW, Calgary, Alberta, T2N 1N4, Canada}
	
	\author{Wolfgang Tittel}
	\altaffiliation{Available for correspondence and requests for materials addressed to W. Tittel (email: \mbox{w.tittel@tudelft.nl).}}
	
	\affiliation{QuTech and Kavli Institute of Nanoscience, Delft University of Technology, 2600 GA Delft, The Netherlands} 
	

	\begin{abstract}
	We design and implement an atomic frequency comb quantum memory for 793 nm wavelength photons using a monolithic cavity based on a thulium-doped Y$_3$Al$_5$O$_{12}$ (Tm:YAG) crystal. Approximate impedance matching results in the absorption of approximately $90\%$ of input photons and a memory efficiency of (27.5$\pm$ 2.7)\% over a 500 MHz bandwidth. The cavity enhancement leads to a significant improvement over the previous efficiency in Tm-doped crystals using a quantum memory protocol. In turn, this allows us for the first time to store and recall quantum states of light in such a memory. Our results demonstrate progress toward efficient and faithful storage of single photon qubits with large time-bandwidth product and multi-mode capacity for quantum networking. 
	\end{abstract}
	
\maketitle

Quantum memories (QM) that can map fast moving quantum states of light reversibly onto matter are an invaluable component of future quantum networks\cite{Divincenzo}. These light-matter interfaces increase the efficiency of complex quantum networking schemes, and allow network tasks to be accomplished over large distances, and via error correction and local processing \cite{wehner2018quantum}. For instance, quantum memories for light play a crucial role in the efficient distribution of entanglement over large distances, thereby securing classical communication by means of quantum key distribution and allowing the loss-less distribution of qubits through quantum teleportation. Common to all these applications is the need for efficient quantum memories that can store many qubits encoded into different modes of light, which is often referred-to as the multiplexing capacity or time-bandwidth product \cite{W2009rev}.

For quantum networking it has been shown that the rate of entanglement generation, or complex multi-photon state generation, scales with the product of memory efficiency and time-bandwidth product \cite{BroadbandWalmsley,MultiPhotonWalmsley}. The intuition behind this follows from that of classical communications: higher efficiency devices allow more equipment to be connected over greater distances before loss takes a toll. Similarly, communication rates scale with the number of temporal and spectral channels, e.g. a memory's multiplexing capacity or time-bandwidth product \cite{sinclair2014spectral, simon2007rate, TBProduct}. Finally, quantum memories for light should feature sufficiently large bandwidths to interface with a diverse range of single photon sources, including spontaneous parametric down-conversion (SPDC), quantum dots, and single molecule emitters \cite{QMforQD,TmErhan,Toninelli:10}. 

Starting with the description of suitable quantum storage protocols around 15 years ago \cite{Moiseev2003,Kraus2006}, cryogenically-cooled rare-earth crystals have rapidly demonstrated their potential as suitable storage materials. In particular, in conjunction with the so-called atomic frequency comb (AFC) protocol \cite{afzelius2009AFC} they have allowed storing non-classical states of light such as single and entangled photons \cite{ANUEFF, Clausen2011, Saglamyurek2011, Maring2017, HUA20191577}. In order to increase efficiencies to values close to 100 \%, there has been a push towards cavity-enhanced light-matter interaction \cite{afzelius2010impedancecavity}. A lot of progress has been reported towards this end \cite{PrSabooni,CavMemGeneva,Zhong1392}, but cavity-enhanced storage of non-classical light remains to be demonstrated, in part due to the limited available memory bandwidth. Here we demonstrate a 500 MHz-broad quantum memory with efficiency up to 27.5\%, as well as high-fidelity storage of heralded single photons over more than 1.5 GHz bandwidth using an impedance-matched cavity. Our findings further support the potential of rare-earth crystals to meet the stringent demands of future quantum networks for memories that allow quantum state storage. 

 \begin{figure*}[t]
 \centering
\includegraphics[width=\textwidth]{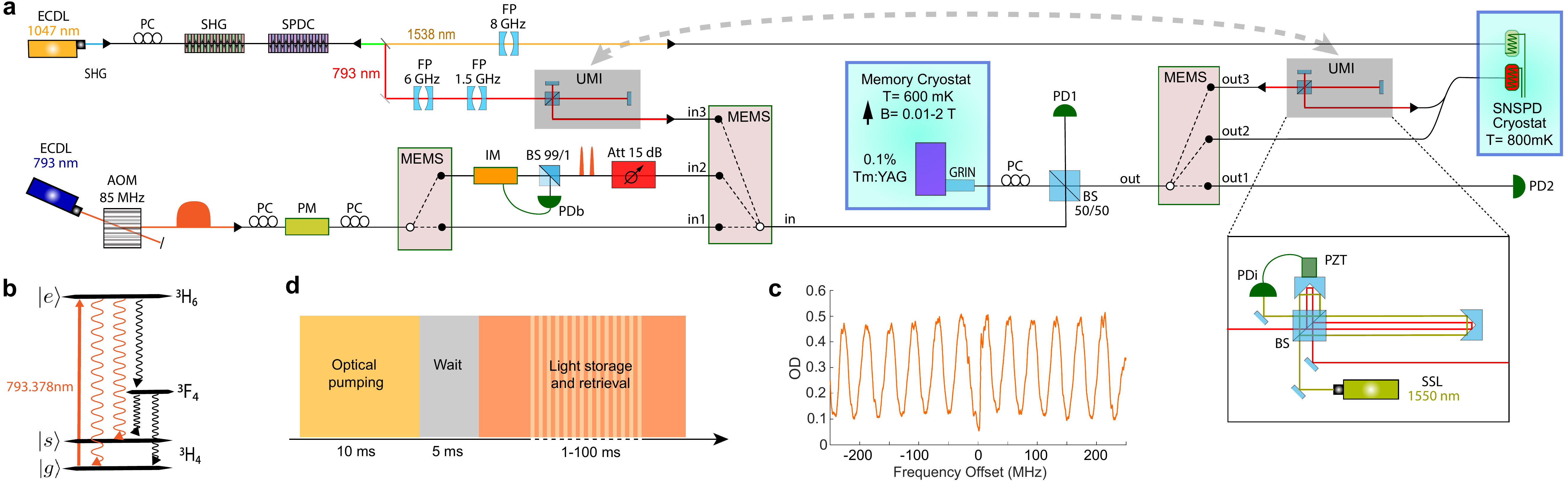}	
\caption {\textbf{(a)} Schematic of experimental setup; FP Fabry Perot Filter Cavity; ECDL External Cavity Diode Laser; SSL Solid State Laser; SHG Second Harmonic Generation; SPDC Spontaneous Parametric Downconversion; FP Fabry-Perot cavity filter; BS Beam-Splitter; PC Polarization Controller; GRIN Gradient Index Lens; UMI Unbalanced Michelson Interferometer; PZT Piezo-electric actuator; Att Optical Attenuator; AOM Acousto-Optic Modulator; PD Photo Diode; MEMS Micro Electro-Mechanical Switches; SNSPD Superconducting Nano-wire Single Photon Detectors; IM Intensity Modulator; PM Phase Modulator. Light from 3 paths was directed  to an input beam splitter for AFC creation and photon storage. After re-emission from the memory, the light was switched between different analyzers. \textbf{(b)} Level structure of Tm:YAG. \textbf{(c)}Example 500 MHz AFC scan of a weak read pulse across the comb. \textbf{(d)} Experimental duty cycle, spectral hole burning time, period of spontaneous emission, and a period for memory use.}
\label{fig:setup}
\end{figure*}
The memory cavity is made of a $l=4$ mm long 0.1\% Tm$:Y_3Al_5O_{12}$ (Tm:YAG) crystal and placed in a cryostat operating at a temperature of 600 mK.  The end facets of the crystal are reflection coated, with reflectivities R$_2$ =99\% on the rear and $R_1=40\%$ on the front side. The reflectivity value for the front facet is chosen to allow for impedance matching at the 793 nm Tm:YAG absorption wavelength by meeting the condition $R_1=R_2e^{(-2\alpha l)}$ with $\alpha$ the average absorption coefficient across the cavity resonance bandwidth \cite{afzelius2010impedancecavity}. These coatings create a planar optical cavity with a free spectral range of 20 GHz, and a finesse of 7. 

The Tm:YAG crystal was cut and mounted such that a magnetic field (aligned along the [001] crystal axis) splits the Zeeman degeneracy for 4 of the 6 crystallographic  sites equally to create a lambda system for optical pumping \cite{Tongning2015}.
Optical access was provided by ferrule-tipped (single mode) fibers and collimation lensing through the planar crystal cavity, and alignment to the fundamental cavity mode was achieved through nano-positioning stages that allow angular steering of the ferrules. All input signals were routed into the memory by a set of MEMS switches and a 50:50 fiber beam-splitter, which also allowed  collection of the reflected signals (see Fig.~\ref{fig:setup}(a)).

A single AFC memory consists of a spectrally periodic distribution of atomic absorption peaks with spacing $\delta$ (see \cite{afzelius2009AFC, Zarkeshian2017,Saglamyurek2011} for more information). Overlapping an optical  signal field with this shaped absorption feature causes the AFC to absorb the input pulse of light. The rare-earth dopants within the crystal are put into an entangled superposition state, primed for re-emission after a pre-set interval $\tau=1/\delta$. We addressed the rare earth ion-doped crystal sample using three paths, as seen in Fig.\ref{fig:setup}(a) -- one for creation of the AFC, and two more for delivering various signals for storage. We sculpt this spectral comb feature through optical pumping on the Tm $^{3}H_{6}$$\rightarrow$$^{3}H_{4}$ transition, shown in Fig.\ref{fig:setup}(b), to drive population with transition frequencies matching the comb troughs into magnetically separated Zeeman level, $|s\rangle$. When the Zeeman level splitting is matched to the desired comb tooth spacing, a spectral grating with finesse (the ratio between peak spacing $\delta$ and peak width) $F_{AFC}\approx$ 2 is prepared without altering the average optical depth. The pumping process, pictured in Fig.\ref{fig:setup}(d), allowed the creation of memory features between 100 MHz-10 GHz bandwidth with tooth spacing from 4-100 MHz. These memories allow storage of 100 ps-10 ns long pulses for times between 10 and 250 ns, limited by pump laser linewidth. An example of one such comb, created in an uncoated region of the crystal, is pictured in Fig.\ref{fig:setup}(c).

\begin{figure}[t]
\includegraphics[width=0.48\textwidth]{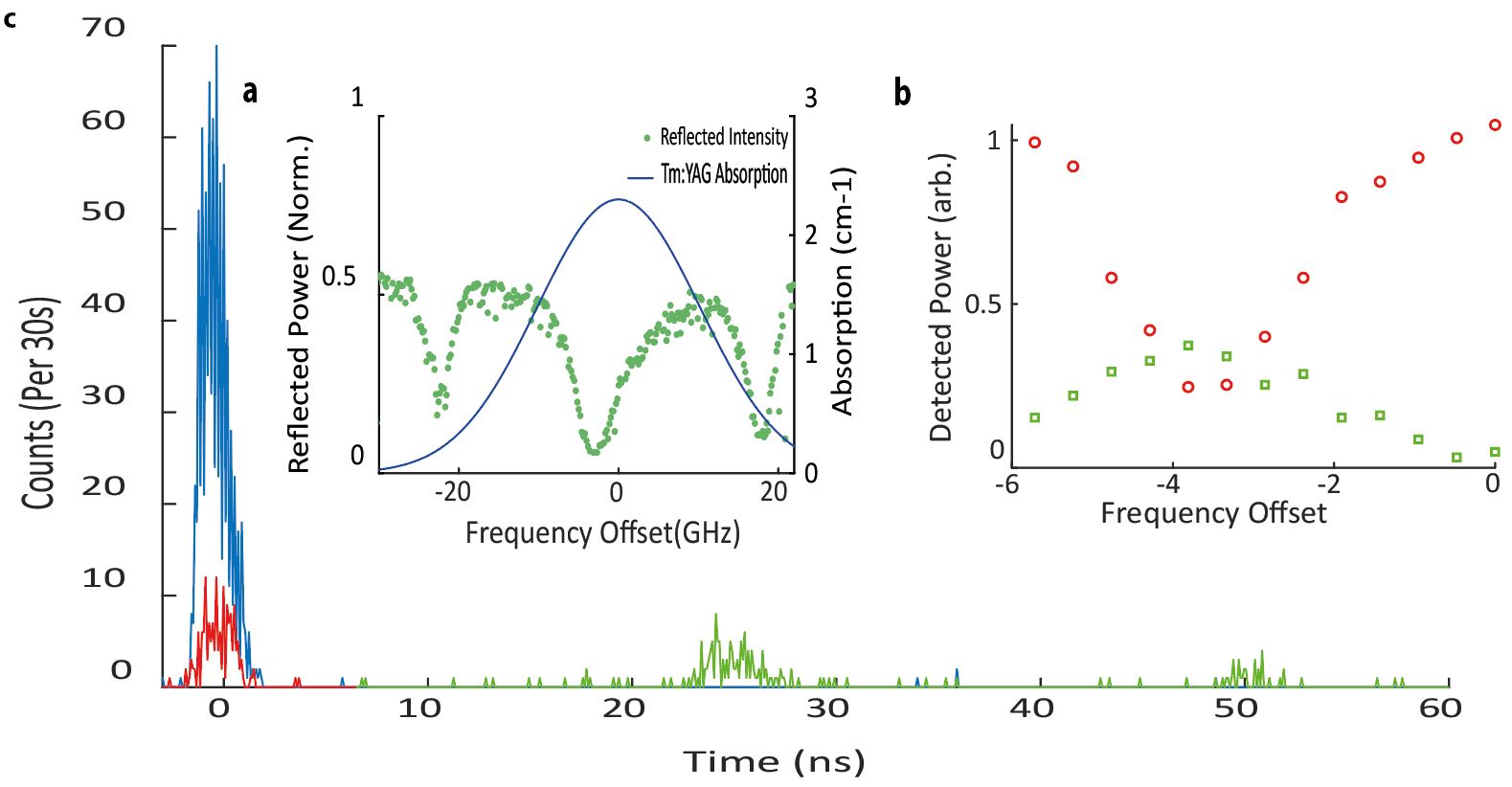}	
\caption {\textbf{(a)} The blue curve shows the absorption of the Tm:YAG crystal without cavity \cite{YAGAbs}, while the green dots show the normalized reflected power obtained by sweeping 4 ns-long pulses of light in frequency across the cavity features. Close-to-perfect impedance matching is obtained at a detuning of -4 GHz. \textbf{(b)} Detected intensity of strong 2 ns-long light pulses after reflection (red dots), and re-emission after 50 ns storage from the cavity (green dots). The maximum re-emission is visible at (approximate) impedance matching. All AFCs were of 500 MHz width. \textbf{(c)} Detection histograms for 25 ns storage of weak coherent pulses. The blue peak is the memory input pulse; the red is the part of the input pulse that is  reflected from the cavity (not stored), the green peaks show recalled pulses after a multiples of the storage time. }
\label{fig2}
\end{figure}

The impedance matching scheme relies on interference between electric field leaking through the front cavity facet after each cavity round trip, and the incoming field initially reflected from the front facet. At the impedance matched condition, these fields create perfectly destructive interference resulting in perfect intake of light by the lossy cavity mode. In our case, the engineered loss (in the form of an AFC) guarantees heightened interaction between light and rare-earth ions. 
In the following we describe initial characterizations of the memory. First, we examined how cavity resonances interact with the Tm:YAG absorption profile. Using an intensity modulator (IM) to carve Gaussian pulses of 4 ns duration at 1000 Hz repetition rate, we slowly swept the laser frequency across the inhomogeneous Tm absorption line centered at 793 nm. Shown in Fig.\ref{fig2}(a), the reflected part of the input pulse was detected and normalized to the input pulse intensity. No signal was detected transmitting through the cavity. On resonance with the cavity mode and $\sim 4$ GHz from thulium line center, more than 90\% of the input energy was absorbed within the rare-earth cavity system. 

Next, we created a 500 MHz-wide AFC and moved its center frequency in 500 MHz increments from the Tm:YAG absorption peak past the cavity resonance. For each detuning, we measured the storage efficiency using laser pulses containing many photons. As expected, the optimum central AFC frequency matched the minimum of the cavity's reflection spectrum. As seen in Fig.~\ref{fig2}(b), the absorption and re-emission of light then becomes more likely than the reflection of the signal light from the cavity, peaking at a system efficiency of 12$\pm$ 1\%. Taking into account 50\% loss of the input-output splitter as well as 6\% coupling and lensing loss, we find a memory efficiency of 27.5$\pm$2.7\%. It decreases to 7\% for a 1.6 GHz wide AFC due to the limited bandwidth of the impedance-matched cavity (see Supplemental Material for details). These values, which are insensitive to the type of input signal (strong or weak laser pulses, or heralded single photons), correspond to a 20 to 30-fold increase of the single-pass efficiency in the same crystal (no cavity), which we estimate to be below 1\%. Note as well that, due to the combination of our sample's small single-pass absorption and the broad bandwidth of our spectral features, we measured no evidence of strong dispersion, which has previously limited bandwidths of similar cavity-based memories \cite{PrSabooni, CavMemGeneva, TianND}.

We used several  light sources to test the memory. First, to create time-bin qubits encoded into temporal modes of attenuated laser pulses, we employed an AOM combined with an IM to tailor a continuous-wave laser beam at 793 nm wavelength. Early and late temporal modes were of 800 ps lengths and separated by 1.4 ns, with spectral and phase control achieved via laser diode grating adjustment and a serrodyne-driven phase modulator.
Second, we prepared quantum-correlated pairs of photons at 1538 and 793 nm wavelength by means of spontaneous parametric down-conversion (SPDC) in a periodically-poled lithium niobate waveguide. After filtering, their spectra were narrowed to 8 GHz and 1.5 GHz, respectively. Third, passing the 793 nm photons before storage through an unbalanced Michelson interferometer, we could furthermore generate heralded time-bin qubits with the same mode separation. All single-photon-level signals were detected using low jitter WSi superconducting nanowire single-photon detects (SNSPDs) \cite{Marsili2013} followed by suitable coincidence electronics. In addition, 
to analyze photons in qubit states, we employed an actively phase-locked Michelson interferometer with 42 cm path-length difference.

\begin{figure}[t]
\includegraphics[width=0.5\textwidth]{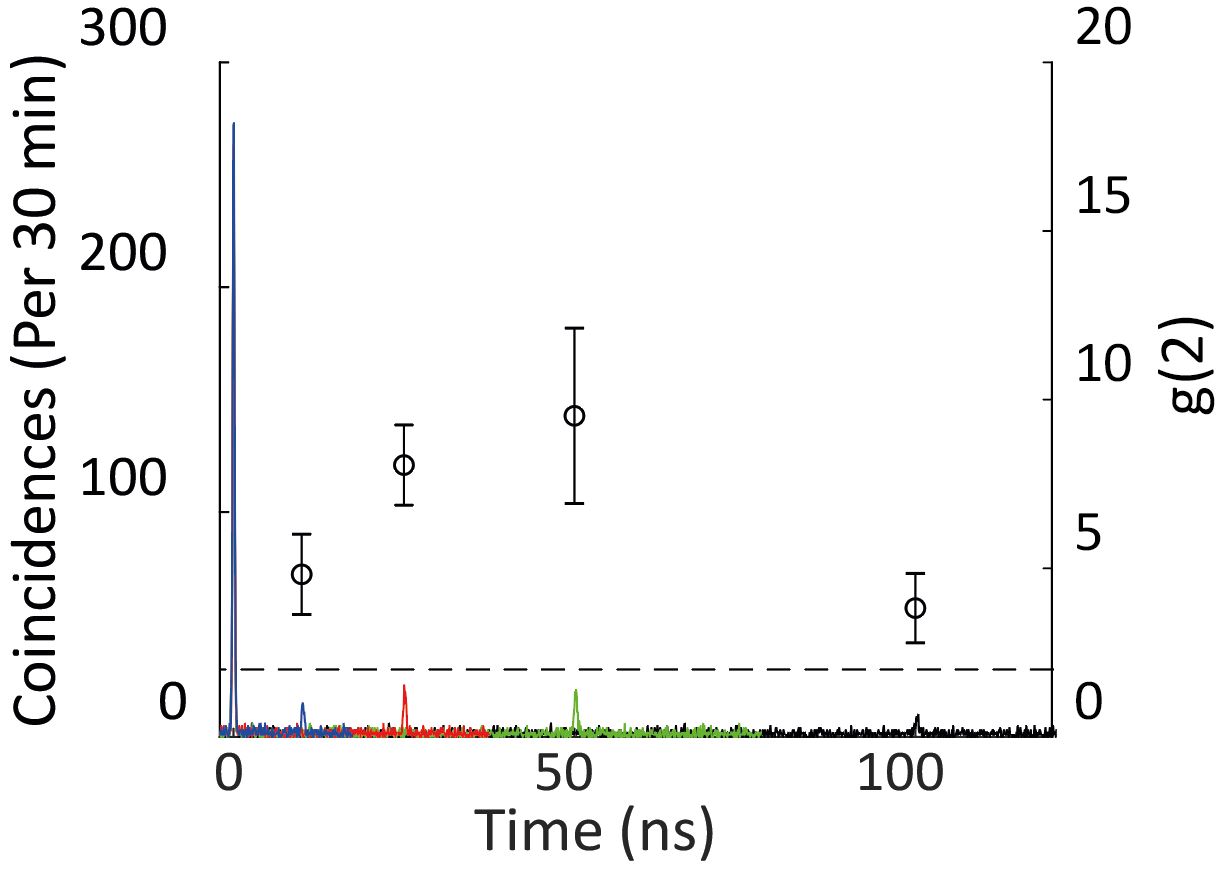}	
\caption {Time-resolved coincidence detections of a 1538 nm heralding photon and a 793 nm photon stored in the cavity memory for various, differently coloured storage times. The peaks visible at each of the set storage times verify that the non-classical correlations created by SPDC persist after storage. The righthand axis depicts the $g^{(2)}$ value for each peak with error bars multiple standard deviations above the classical limit (dashed line).}
\label{fig3}
\end{figure}

To study how well our memory stores non-classical light, we first used photon pairs and measured cross correlations between heralding 1538 nm photons and 793 nm photons. At maximum pump laser power with no memory in place, we found a cross-correlation coefficient of $g^{(2)}=61.8 \pm 3.8$. For values of $g^{(2)} > 2$, the correlations violate the Cauchy-Schwartz inequality, implying that they are non-classical in nature \cite{kuzmich2003generation}. Adding the memory for the 793 nm photons, we measured $g^{(2)}=9.1 \pm 1.2$ after 25 ns storage, and we found Cauchy Schwartz violations after as much as 100 ns (see Fig.~\ref{fig3}). This shows the preservation of quantum correlations during storage in the cavity memory, and hence establishes that the cavity-enhanced memory as a quantum memory for light.	
	
\begin{figure}[t]
\includegraphics[width=0.5\textwidth]{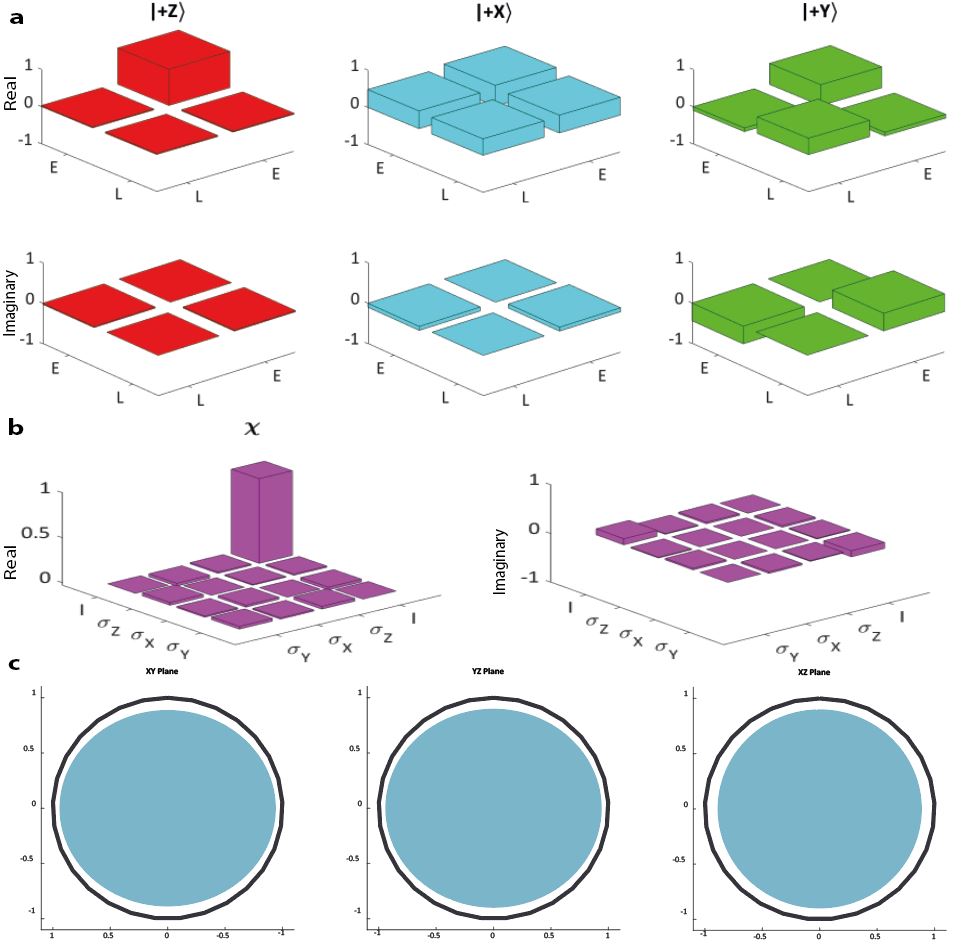}	
\caption {\textbf{a.} Density matrices of three non-orthogonal qubit states after re-emission form the memory. \textbf{b.} Quantum process matrix $\chi$. \textbf{c.} Cross sections through the Bloch sphere depicting the average over all possible output states. Black circles denote the output expected from a perfect memory, and the filled ovals depict qubits after the non-ideal storage process \cite{RempeAtomMem}.}
\label{fig4}
\end{figure}

Next, we created and stored various time-bin qubits encoded into attenuated laser pulses with a mean photon number of $\mu=0.7$. For these measurements we set the storage time to 25 ns.
For Z basis states, $\ket{Z+}\equiv\ket{e}$ and $\ket{Z-}\equiv\ket{l}$, we found a mean fidelity of $\mathcal{F}_{z}=97.6\pm 0.02\%$. Furthermore, characterization of $\ket{X_{\pm}}\equiv\frac{\ket{e} \pm \ket{l}}{\sqrt{2}}$ and $\ket{Y_{\pm}} \equiv \frac{\ket{e} \pm i\ket{l}}{\sqrt{2}}$ yielded $\mathcal{F}_{x,y}=93.7\pm 0.1\%$, resulting in a memory fidelity averaged over all six states of $ \mathcal{F}= \frac{1}{3}\mathcal{F}_{z}+ \frac{2}{3}\mathcal{F}_{x,y}=95 \pm 0.1\%$ (see Fig.~\ref{fig4}, the Supplemental Material, and \cite{TmErhan} for more details). Taking into account the recall efficiency of 7\% and the mean photon number of $\mu=0.7$, all fidelities significantly exceed the upper bound of $\mathcal{F}(\mu, \eta)=80.3 \%$, established conservatively for classical memories under the assumption  of intercept-resend attacks \cite{Membound, RempeAtomMem, Gundogan}, as well as that imposed by the optimal universal quantum cloning machine \cite{Cloner}.

Finally, we repeated these measurements after replacing the source of attenuated laser pulses by heralded single photons. But instead of using interferometers to analyze the qubits after storage in the X- and Y-basis, we configured the memory for double-comb storage with storage-time separation matching the qubit time-bin separation. This method allows for a convenient analysis of time-bin qubit states in the superposition bases \cite{NdHugue}. The series of measurements resulted in a heralded single-photon qubit fidelity of $\mathcal{F}=85 \pm 0.02 \%$(See Supplemental Material), again exceeding the thresholds imposed by classical storage and quantum cloning. It also allowed us to establish the density matrices of recalled qubits for various inputs by means of quantum state tomography (see Fig.~\ref{fig4}(a)), and in turn to perform quantum process tomography \cite{ProcessTom, ProcessTom2}. The resulting storage process matrix, $\chi$, depicted in Fig.~\ref{fig4}(b), complete describes the mapping between arbitrary input and output qubit states\cite{RempeAtomMem}. As expected, the dominant term describes the identity operation, but some small imperfections with magnitudes $\leq 0.1$ are also visible. 

Another way of presenting these imperfections is to look at the ensemble of output states averaged over all possible input states. This results in a deformed Bloch sphere, as shown in Fig.~\ref{fig4}(c). But note that this deformation as well as the unexpected elements in the process matrix in Fig.~\ref{fig4}(b) provide the upper bound to imperfections in the storage process. Indeed, they are mostly caused by imperfect state preparation and measurement, rather than by an imperfect memory. For example, non-ideal PM drive-pulse duration and timing cause small frequency shifts between temporal input modes, making them distinguishable and therefore reducing the quality of the interference required for analysis. Furthermore, imbalances of the beamsplitters used to split and couple temporal modes in our analysis interferometer also lower the measured fidelities.

To summarize, we have demonstrated that solid-state quantum memory based on ensembles of rare-earth ions and cavity-enhanced light-matter interaction in a monolithic and fibre-coupled cavity allows storing quantum states of light. Our memory operates in the domain of pre-set storage times and allows feed-forward based mode mapping using external frequency shifters \cite{sinclair2014spectral}. It is capable of storing broadband quantum light with a fidelity of at least 95\%, an efficiency $\eta$ of up to 27.5\%, and a time-bandwidth product TB of up to 100 ns *1.5 GHz=150, where 100 ns is the longest time after which we observed a violation of the Cauchy-Schwartz inequality. We note that the factor $\eta * TB=1050$ (with $\eta=7$\%), which is comparable to that obtained in previous demonstrations of rare-earth-ensemble-based storage of attenuated laser pulses \cite{Holzapfel2019}.

To further improve key properties, several modifications are required. First, to reach an efficiency close to 100\%, the finesse of the AFC has to be increased beyond its current value of two. As this is only possible if the total AFC width is smaller than the ground-state splitting of the  rare-earth ion, in this case 200 MHz/T, this limits the available bandwidth per spectral channel, implying the need for suitably adapted sources of quantum light. However, the time-bandwidth product can remain high as many spectral channels can be created in parallel \cite{sinclair2014spectral}. Furthermore, fibre coupling loss must be reduced through better mode matching. Second, to increase the storage time to a few hundred $\mu$sec, enough for an elementary link in a quantum repeater architecture of around 100 km \cite{sinclair2014spectral}, materials with improved coherence have to be employed. Possibilities include Tm:Y$_3$Ga$_5$O$_{12}$  for which optical coherence times of 490$\mu$sec have been reported \cite{thiel2014Tm:YGG}, or materials featuring narrow ground-state transitions \cite{Tongning2015}.

\subsection*{Acknowledgements}  
The authors would like to thank Neil Sinclair for discussions, and Jay Mckisaac and Nik Hamilton for design assistance and manufacturing of cryogenic components. We acknowledge funding through Alberta Innovates Technology Futures (AITF), the National Science and Engineering Research Council of Canada (NSERC),the Netherlands Organization for Scientific Research (NWO), and the European Union’s Horizon 2020 research and innovation program under grant agreement No 820445 and project name Quantum Internet Alliance. Furthermore, WT acknowledges funding as a Senior Fellow of the Canadian Institute for Advanced Research (CIFAR).

\section*{References}


%
	
\end{document}


\captionsetup[figure]{labelfont={bf},name={SIFig.},labelsep=period}
 	\title{Supplemental Material: Improved Light-Matter Interaction for storage of quantum states of light in a Thulium-doped crystal\\}
	
	\author{Jacob H. Davidson}
	\affiliation{Institute for Quantum Science and Technology, and Department of Physics \& Astronomy, University of Calgary, 2500 University Drive NW, Calgary, Alberta, T2N 1N4, Canada}
	\affiliation{QuTech and Kavli Institute of Nanoscience, Delft University of Technology, 2600 GA Delft, The Netherlands}
		\author{Pascal Lefebvre}
	\affiliation{Institute for Quantum Science and Technology, and Department of Physics \& Astronomy, University of Calgary, 2500 University Drive NW, Calgary, Alberta, T2N 1N4, Canada}
		\author{Jun Zhang}
	\affiliation{Institute for Quantum Science and Technology, and Department of Physics \& Astronomy, University of Calgary, 2500 University Drive NW, Calgary, Alberta, T2N 1N4, Canada}
	\author{Daniel Oblak}
	\affiliation{Institute for Quantum Science and Technology, and Department of Physics \& Astronomy, University of Calgary, 2500 University Drive NW, Calgary, Alberta, T2N 1N4, Canada}
	
	\author{Wolfgang Tittel}
	\affiliation{QuTech and Kavli Institute of Nanoscience, Delft University of Technology, 2600 GA Delft, The Netherlands} 
	
 \maketitle

 \begin{figure*}[h]
 \centering
\includegraphics[width=0.5\textwidth]{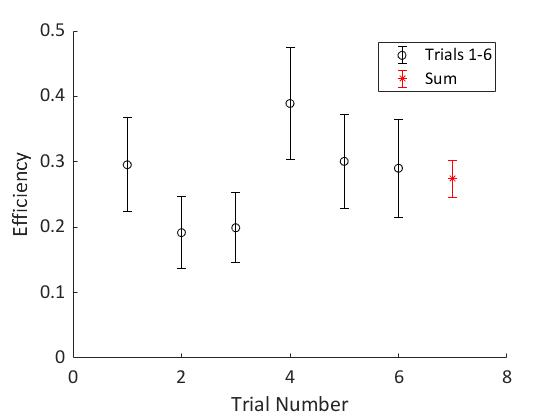}	
\caption {Memory Efficiency over a number of trials with the optimal polarization setting. Each trial is the summed results for memory efficiency from 30 seconds of data collection for attenuated classical light at the single photon level. The final point shown in red is the sum of all trials which yields the efficiency listed in the main text.}
\label{SIFig2.}
\end{figure*}

 \begin{figure*}[h]
 \centering
\includegraphics[width=0.5\textwidth]{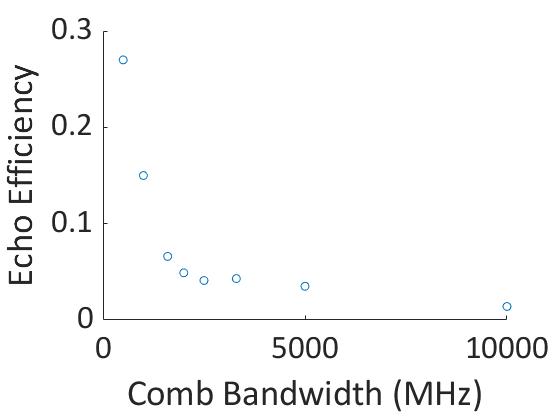}	
\caption {Scaling of Maximum AFC memory efficiency with bandwidth of the created comb feature. All combs were created centered on the cavity resonance -4 GHz from the absorption line center. Each point is the measured memory efficiency after 25ns of storage using attenuated light pulses with bandwidth matching the comb feature.}
\label{SIFig1.}
\end{figure*}

 \begin{figure*}[h]
 \centering
\includegraphics[width=0.75\textwidth]{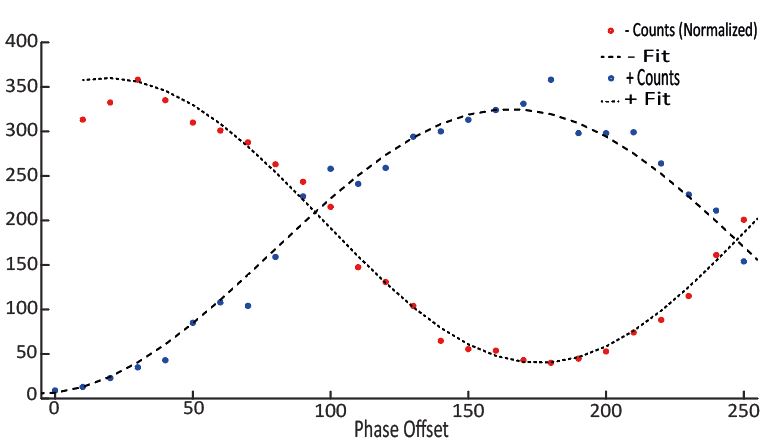}	
\caption {Interference curves generated by $|+\rangle$ and $|-\rangle$ time bin qubits created from weak coherent pulses after storage in the memory.}
\label{SIFig6.}
\end{figure*}

 \begin{figure*}[h]
 \centering
\includegraphics[width=0.75\textwidth]{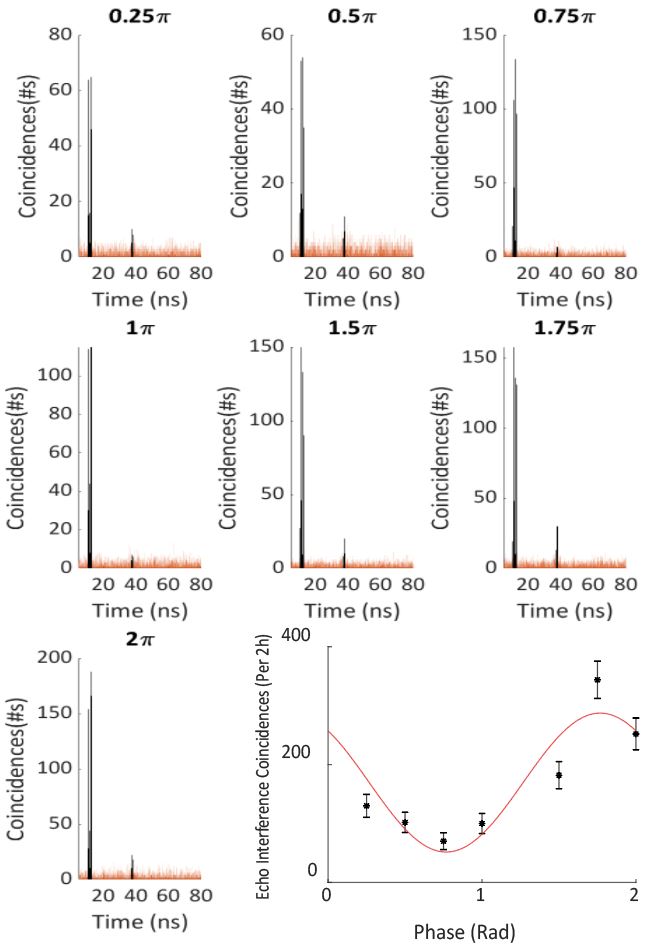}	
\caption {Each plot shows coincidences histograms generated by detection of time bin qubits in the $|+\rangle$ state from the SPDC source after storage for differently phased double combs. The central interference echo peak is highlighted in each plot. \textbf{(Inset)} Normalized Coincidence counts after projection of heralded single-photon qubits in state $\ket{+X}$ after re-emission onto $\ket{\psi}=(\ket{e}+exp\{i\varphi\}\ket{l})/\sqrt{2}$ using differently phased double comb features.
}
\label{SIFig5.}
\end{figure*}